\documentclass{aa}  

\usepackage{graphicx}
\usepackage{txfonts}
\usepackage{enumitem}
\usepackage{xcolor}
\usepackage{hyperref}
\usepackage{siunitx}
\usepackage{amsmath,amssymb}
\usepackage{textcomp}
\usepackage{gensymb}

\newcommand{\orcid}[1]{\href{https://orcid.org/#1}{\includegraphics[width=8pt]{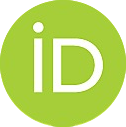}}}

\newcommand{\project}[1]{\textsl{#1}}
\newcommand{\apogee}{\project{\textsc{apogee}}}
\newcommand{\feh}{\mbox{$\rm [Fe/H]$}}

\newcommand{\mgfe}{\mbox{$\rm [Mg/Fe]$}}
\newcommand{\logg}{\mbox{$\log (\mathrm{g})$}}

\defcitealias{Portail_dyn_2017}{P17}
\begin{document} 

   \title{The Milky Way's middle-aged inner ring}

  \author{Shola M. Wylie \inst{1} \orcid{0000-0001-9116-6767} \thanks{swylie@mpe.mpg.de} \and Jonathan P. Clarke \inst{1} \orcid{0000-0002-2243-178X} \and Ortwin E. Gerhard \inst{1} \orcid{0000-0003-3333-0033}}

  \institute{Max-Planck-Institut fur Extraterrestrische Physik, Gießenbachstraße, D-85748 Garching, Germany}

  \abstract
   {}
   {We investigate the metallicity, age, and orbital anatomy of the inner Milky Way, specifically focussing on the outer bar region.}
   {We integrated a sample of $\apogee$ DR16 inner Galaxy stars in a state of the art bar-bulge potential with a slow pattern speed and investigated the link between the resulting orbits and their $\feh$ and ages. By superimposing the orbits, we built density, $\feh$, and age maps of the inner Milky Way, which we divided further using the orbital parameters eccentricity, $|\mathrm{X}_{\mathrm{max}}|$, and $|\mathrm{Z}_{\mathrm{max}}|$.}
   {We find that at low heights from the Galactic plane, the Galactic bar gradually transitions into a radially thick, vertically thin, elongated inner ring with average solar $\feh$. This inner ring is mainly composed of stars with AstroNN ages between 4 and 9 Gyr with a peak in age between 6 and 8 Gyr, making the average age of the ring  ${\sim}6$ Gyr. The vertical thickness of the ring decreases markedly towards younger ages. We also find very large L4 Lagrange orbits that have average solar to super-solar metallicities and intermediate ages. Lastly, we confirm a clear X-shape in the $\feh$ and density distributions at large Galactic heights.}
   {The orbital structure obtained for the $\apogee$ stars reveals that the Milky Way hosts an inner ring-like structure between the planar bar and corotation. This structure is on average metal rich, intermediately aged, and enhances the horizontal metallicity gradient along the bar's major axis.}
   
   
   \date{Received-; accepted -}

   \maketitle
%
%-------------------------------------------------------------------

\section{Introduction}\label{intro}
From previous Galactic bulge studies, we know that the stars in the outer regions of the Galactic bar are on average more metal rich and younger in comparison to the stars in the central bulge \citep{Bovy_2019, Hasselquist_2020, Queiroz_2020b}. This difference leads to a pronounced horizontal metallicity gradient along the bar's major axis \citep{Wylie_2021}. The Milky Way (MW) is not unique in this structure; bars with ends that are more metal rich and/or younger than their central regions have been observed in a few other galaxies as well \citep{Seidel_2016}, including M31 \citep{Gajda_2021}.

This gradient structure has several possible origins, one of which is that the bulge and bar were formed from dynamical instabilities of coexisting discs with differing scale lengths and metallicities \citep{Fragkoudi_2018, Wylie_2021}. In this scenario, the most metal-poor discs dominate in the very central regions due to their shorter scale lengths, while the more metal-rich discs dominate in the outer regions, resulting in a positive horizontal metallicity gradient. However the MW's horizontal gradient is likely too steep to be fully explained by this scenario, suggesting that additional mechanisms are at work. 
To investigate this further, we built density, $\feh$, and age orbital maps for a sample of inner Galaxy stars from  $\apogee$\footnote{Apache Point Observatory Galactic Evolution Experiment} DR16, for which these parameters along with their positional and 3D kinematic information are available from the ASPCAP\footnote{APOGEE Stellar Parameter and Chemical Abundance Pipeline}, AstroNN, and Gaia DR2 catalogues \citep{Garc_a_P_rez_2016, gaia_colab_2018, Leung_2019, Mackereth_2019}. We integrated the orbits of these stars in a realistic MW bar-bulge potential from \citet[][hereafter P17]{Portail_dyn_2017} which was fit to MW star count data derived from the VVV, UKIDSS, and 2MASS surveys\footnote{Vista Variables in the Via Lactea, UKIRT Infrared Deep Sky Survey, Two Micron All Sky Survey} \citep[][]{Saito_2012, Lucas_2008, Skrutskie_2006} by \citet{Wegg_2013} and \citet{Wegg_2015}, and to kinematic data from the BRAVA, ARGOS, and OGLE\footnote{Bulge Radial Velocity Assay, Abundances and Radial
velocity Galactic Origins Survey, Optical Gravitational Lensing Experiment} surveys \citep{Kunder_2012, Ness_2013_IV, Rattenbury_2007}. The favoured model from P17 had a slow pattern speed of $\Omega_b =\!39 \pm 3.5 \, \mathrm{km} \, \mathrm{s}^{-1} \, \mathrm{kpc}^{-1}$ which puts the corotation radius of the bar at slightly over 6 kpc. More recent dynamical studies of both the stellar kinematics in the bar and resonant stars in the solar neighbourhood have found similar or slightly slower values of $\Omega_b$ \citep[e.g.][]{Bovy_2019, Binney_2020, Chiba_2021, Li_2021, Clarke_2021}.

As we later see, a mechanism to enhance the horizontal gradient in the bar is an inner ring between 4-6 kpc along the bar's major axis. This has also been seen in some galaxy models in cosmological simulations \citep{Fragkoudi_2020}. Inner rings are observed in a substantial fraction of barred disc galaxies \citep{Kormendy_1979, Buta_1996, Comeron_2014} and tend to be preferentially aligned with the bar. Inner rings observed at optical wavelengths generally contain star-forming regions \citep{Buta_1995} and are thought to form from the collection of gas around the bar's 4:1 resonance \citep{Schwarz_1984, Buta_1996}. Passive rings which no longer form stars can be thicker than active rings, but they are generally seen in earlier type galaxies \citep[$\geq$ Sab,][]{comeron_2013}.

This paper is structured as follows: In Sect.\ref{methods} we explain, in more detail, the sample of stars we used, the potential we integrated them in, and the assumptions we made. We also compare the distributions of the $\apogee$ stars in heliocentric velocity space to predictions from the dynamical model used to generate the potential to check that they are consistent. In Sect.\ref{results} we show density maps for the model and the $\apogee$ stellar orbits, and $\feh$ and age maps for the $\apogee$ orbits selected using spatial symmetry and orbital eccentricity criteria. We end in Sect.\ref{conclusions} with a discussion of our results and our conclusions.

%%%%%%%%%%%%%%%%%%%%%%%%%%%%%%%%%%%%%%%%%%%%%%%%%%%%%%%%%%%%%%%%%%%%%%%

\section{Data and methods}\label{methods}
\begin{figure}
\centering
\includegraphics{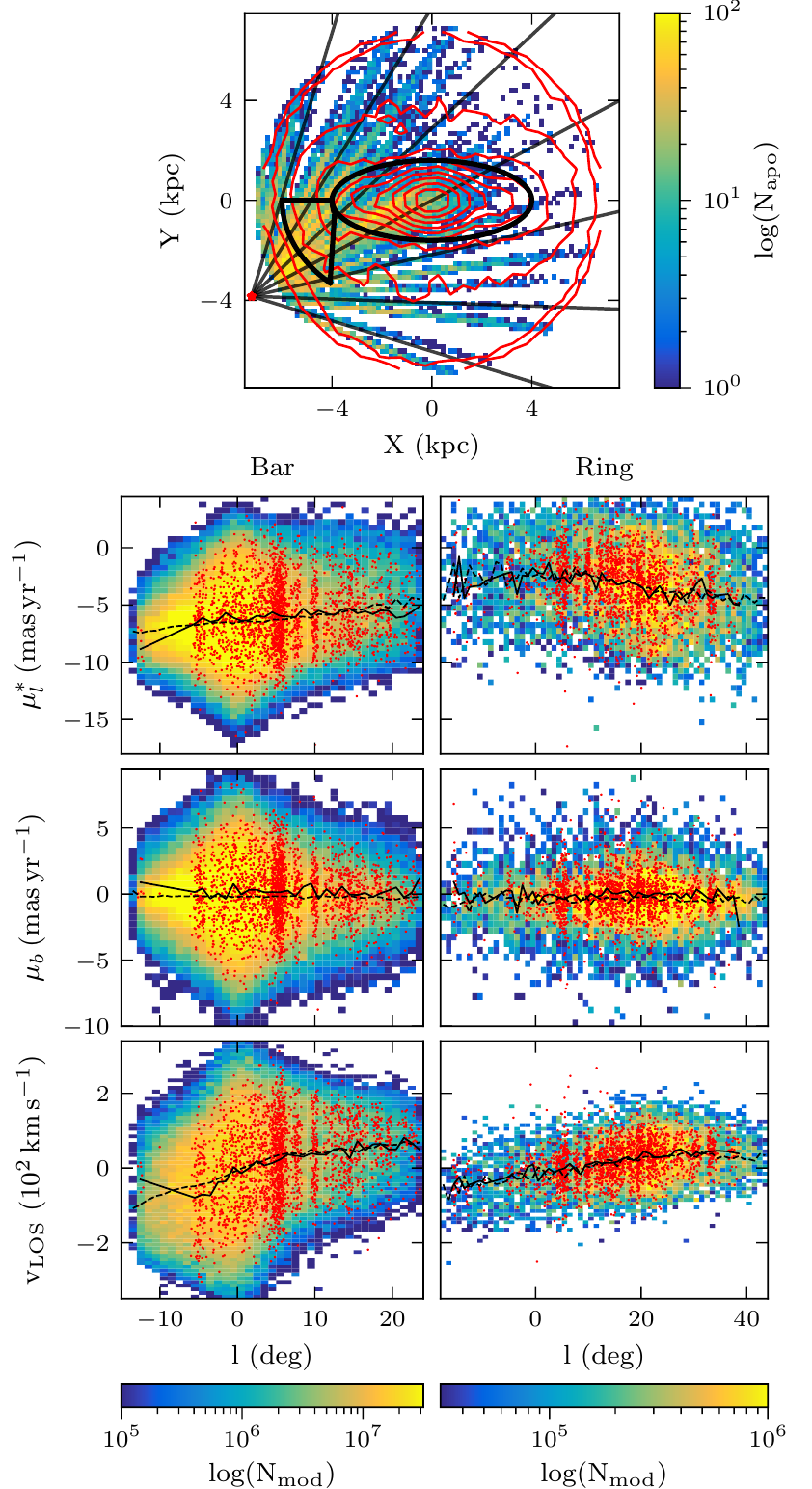}
\caption{Comparisons of the model particles and $\apogee$ stars. Top: Bar frame density distributions of the model particles (red contours) and $\apogee$ stars, both restricted to $|b|<13 \degree$ and $\mathrm{R}_{\mathrm{GC}}<7$ kpc. The straight grey lines mark sight lines at longitudes of $0\degree$, $\pm 15\degree$, $\pm 30\degree$, and $\pm 45\degree$ and the red star shows the Sun's position. Bottom: Comparison of the heliocentric velocity distributions of model particles (2D histograms) and $\apogee$ stars (red points) with $|\mathrm{b}|<3\degree$ in the bar region (left column, black ellipse in the top plot) and in a nearby section of a ring around the bar (right column, black arc in the top plot). Dashed and solid lines give the running means of the model and $\apogee$ distributions, respectively.}
\label{fig:mod_data_helio}
\end{figure}

\begin{figure*}
\centering
\includegraphics{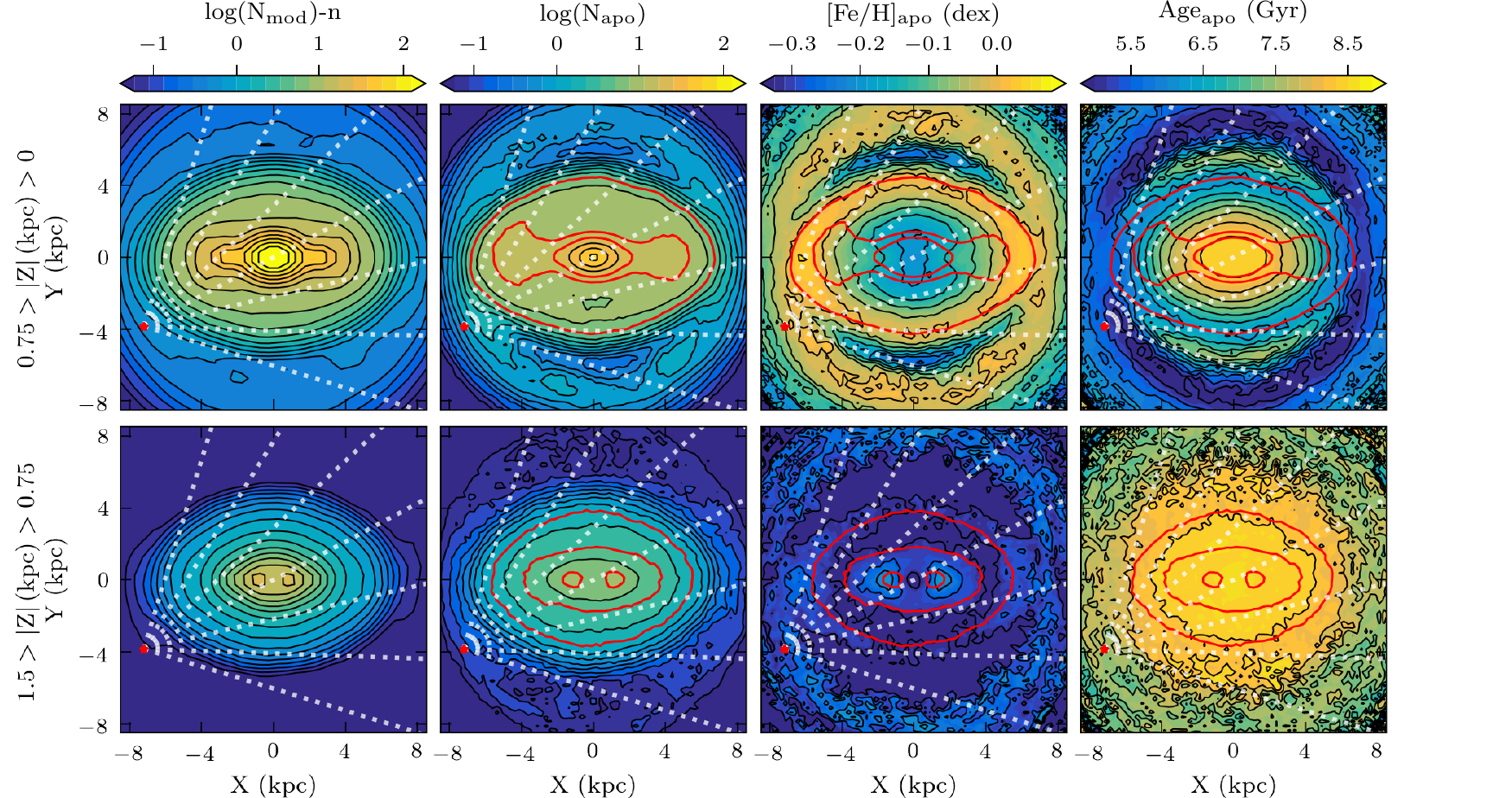}
\caption{Orbital maps of the model and the $\apogee$ stars at different heights above the plane, using only Y-symmetric orbits with $-1\leq \feh \, (\mathrm{dex})<0.5$. Left two columns: Density maps built from the superposition of the model (left) and $\apogee$ stellar (right) orbits. The model density map has been normalised to a similar scale as the $\apogee$ density map using the factor n. Right two columns: Mean $\feh$ (left) and AstroNN age (right) $\apogee$ orbital maps. The top panels show the orbital maps of the stars closer to the plane. In all plots, the red star marks the position of the Sun, while the white dashed lines mark sight lines at the following longitudes: $0\degree$, $\pm 15\degree$, $\pm 30\degree$, and $\pm 45\degree$. The red contours in the $\apogee$ maps show the specific density levels that were chosen to highlight important features and to guide the eye. The $\apogee$ SF has not been corrected for.}
\label{fig:maps_nc}
\end{figure*}

In this work we used stars from the $\apogee$ DR16 catalogue \citep{Majewski2016}. For each star, we obtained its $\feh$ and line-of-sight velocity from the ASPCAP pipeline \citep{Garc_a_P_rez_2016, Holtzman_2018, jnsson2020apogee} and its RA and Dec proper motions from the Gaia DR2 Catalogue \citep{gaia_colab_2018}. Additionally, we obtained a spectrophotometric distance and age for each star from the AstroNN catalogue \citep{Leung_2019, Mackereth_2019}. We restricted the sample of stars to those in the $\apogee$ main sample (EXTRATARG flag$=0$) with valid ASPCAP $\feh$, effective temperatures ($\mathrm{T}_{\rm eff}$), surface gravities ($\logg$), and $\mgfe$. As quality cuts, we also required S/N$>\!60$, ASPCAP $\mathrm{T}_{\rm eff}\!>\!3200$ K, AstroNN distance errors less than $20\%$, $0\leq \mathrm{AstroNN \, \, age} \, (\mathrm{Gyr})<12$, and no $\mathrm{Star}\_\mathrm{Bad}$ flag set (23rd bit of ASPCAPFLAG = 0). For the analysis of this work requiring the AstroNN ages, we restricted the stars to only those with AstroNN $\logg$ errors $<$0.2 dex to remove dwarf stars and AstroNN $\feh>-0.5$ dex as recommend by \citet{Bovy_2019}. Lastly, to focus on the inner MW, we made a spatial cut, requiring the stars to have cylindrical distances from the Galactic centre (GC) $\mathrm{R}_{\mathrm{GC}}$ $<7$ kpc and absolute latitudes ($|\mathrm{b}|$) $<13\degree$. This leaves $32,536$ stars in our sample (26,416 in the age sample). We found that when we restricted our sample further by requiring the stars to have RA and Dec proper motion errors less than $0.5 \, \mathrm{mas} \,\mathrm{yr}^{-1}$ and distance errors less than $10\%$ (removes ${\sim}9,000$ stars), our main results essentially remained unchanged.

We integrated the orbits of our $\apogee$ sample in the rotating potential of one of the \citetalias{Portail_dyn_2017} dynamical bar models. P17 adapted these models using the made-to-measure method, such that they fitted the red clump density from the VVV, UKIDSS, and 2MASS surveys and the stellar kinematics from the BRAVA, OGLE, and ARGOS surveys. They assumed a bar angle $\phi_\odot=28\degree$ and the Sun's distance from the GC $R_0=8.2$ kpc. The dark matter potential was also adapted during these fits. For the results of the work here, we used the P17 model with a pattern speed $\Omega_b\!=\!37.5 \, \mathrm{km} \, \mathrm{s}^{-1} \, \mathrm{kpc}^{-1}$, their central disc mass $M_c=2 \times10^9$ M\textsubscript{\(\odot\)}, and mass to red clump star number of 1000 M\textsubscript{\(\odot\),} as it has provided a good match to both the bulge proper motions \citep[][]{Clarke_2019} and inner Galaxy gas flows \citep{Li_2021}. As a test, we also integrated the $\apogee$ stars in \citetalias{Portail_dyn_2017} models with $\Omega_b\!=\!35 \, \mathrm{km} \, \mathrm{s}^{-1} \, \mathrm{kpc}^{-1}$ and $\Omega_b\!=\!40 \, \mathrm{km} \, \mathrm{s}^{-1} \, \mathrm{kpc}^{-1}$, finding that a change in the pattern speed results in quantitative changes to the shapes of the structures we found; however, the main results of the paper remain unchanged. 

For the orbit integration, we used the inbuilt leap frog integration algorithm (drift-kick-drift with an adaptive time step) of the NMAGIC code \citep{delorenzi_2007}, integrating our $\apogee$ sample for 2 Gyr and saving each orbit's trajectory every 1 Myr. When transforming the $\apogee$ stars to the bar frame and the model to the heliocentric frame, we took $\phi_\odot=28\degree$ and the Sun's position and 3D velocities to be ($\mathrm{R}_{\mathrm{GC}}$, $\mathrm{Z}_\odot$)=(8.178 kpc, 20.8 pc) and ($\mathrm{V}_{\phi, \odot}$, $\mathrm{V}_{\mathrm{r}, \odot}$, $\mathrm{V}_{\mathrm{z}, \odot}$) = ($248.54 \mathrm{km} \, \mathrm{s}^{-1}$, $11.1 \mathrm{km} \, \mathrm{s}^{-1}$, $7.25 \mathrm{km} \, \mathrm{s}^{-1}$), respectively \citep{Schonrich_2010, Bennett_2019, Gravity_Collab_2019, Reid_2020}. We define the bar frame such that X and Y are along the major and minor axes of the bar, respectively, with the Sun at negative X and Y (see red star in the top plot of Fig.\ref{fig:mod_data_helio}). As an additional test, we transformed the $\apogee$ stars to the bar frame assuming $\phi_\odot=25\degree$ and reran the orbit integration. We found that while there were minor differences in the details, there were no major differences in our results.

In the top plot of Fig.\ref{fig:mod_data_helio}, we show the bar frame density distributions of the model particles and all $\apogee$ stars used in this work. The $\apogee$ survey's spatial selection function (SF) is clearly visible. We therefore checked that the heliocentric velocity distributions, that is the longitude and latitude proper motion ($\mu_{l}^{*}$ and $\mu_{b}$) and line-of-sight velocity ($v_{LOS}$) distributions, are consistent for $\apogee$ stars and model particles selected from similar spatial regions in the bar frame. The bottom plots of Fig.\ref{fig:mod_data_helio} show this comparison for stars with $|\mathrm{b}|<3\degree$ in the bar and an adjacent section of a ring around the bar, respectively. Here we define the bar as an ellipse orientated along X in the bar frame, with a major axis length and an axis ratio of 4 kpc and 0.4, respectively. The ring section is defined as the region between two ellipses with major axes of 4 and 6 kpc and with $\mathrm{X}$<$-4$ kpc and $\mathrm{Y}$<$0$ kpc (see the top plot of Fig.\ref{fig:mod_data_helio}). This region is well populated by $\apogee$ stars. Fig.\ref{fig:mod_data_helio} shows that the model particles and $\apogee$ stars generally overlap in heliocentric phase space when selected from similar bar frame spatial regions, and that their mean velocities at each longitude, when well populated, also generally agree (also see Fig.~19 of P17 comparing $\apogee$ DR12 with a similar model). The model's standard deviations of the three heliocentric velocities at each longitude agree with those of $\apogee$ in the bar region, and they are hotter by about $20\%$ in the ring region. We note that the model was not fit to kinematic data in the planar bar region.

%%%%%%%%%%%%%%%%%%%%%%%%%%%%%%%%%%%%%%%%%%%%%%%%%%%%%%%%%%%%%%%%%%%%%%%

\section{Results}\label{results}
\begin{figure*}
\centering
\includegraphics{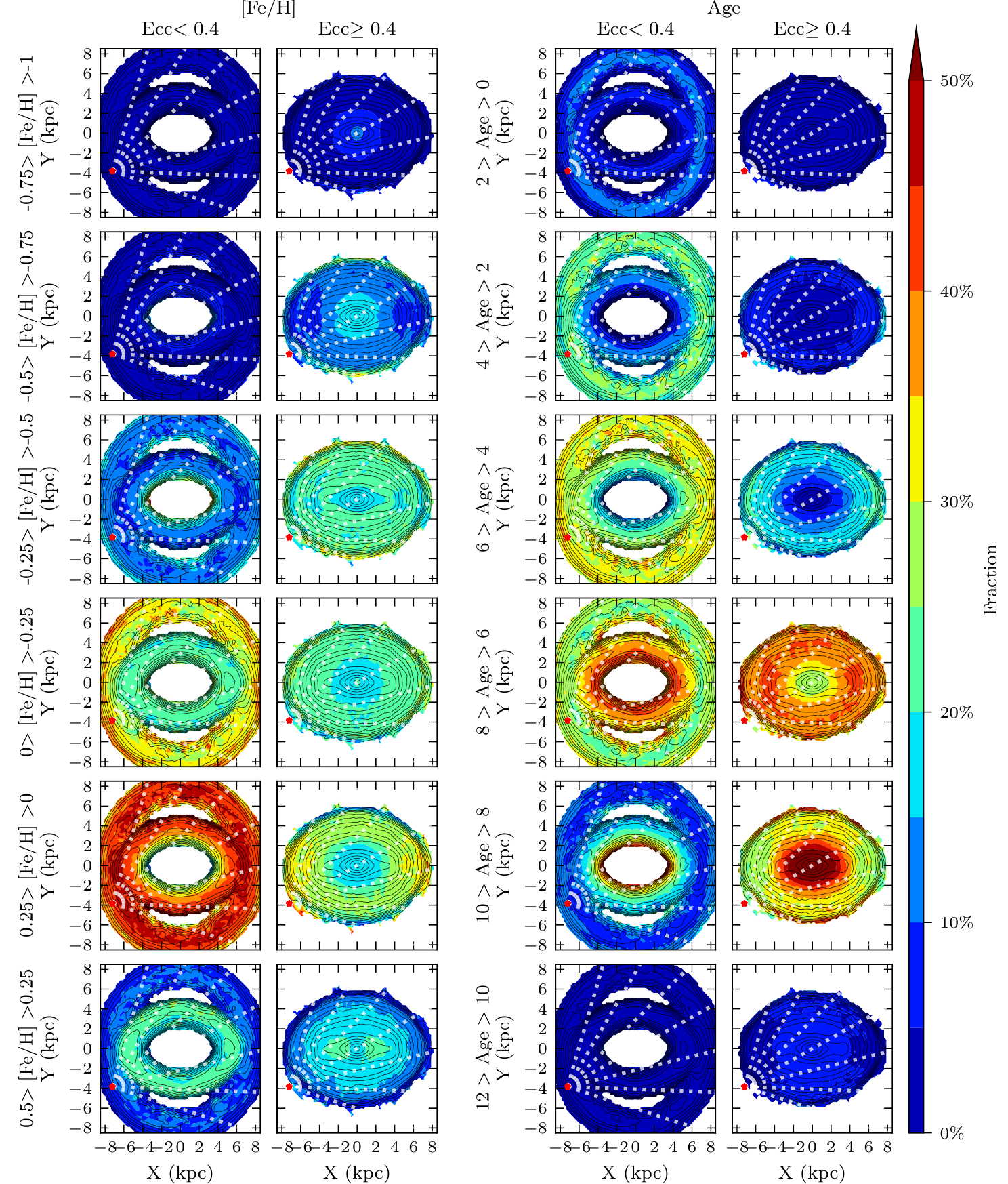}
\caption{Fraction of stars (orbital time steps) along Y-symmetric $\apogee$ orbits with $|\mathrm{Z}|<0.75$ kpc, in different $\feh$ and age bins that constitute the disc, ring, and bar. The black curves show the orbital density maps built from all disc and ring stars ($\mathrm{Ecc}<0.4$) and all bar stars ($\mathrm{Ecc}\geq0.4$). The fraction at each position of stars in each $\feh$ and age bin (rows) composing each structure is shown by the colour. For other plot details see Fig.\ref{fig:maps_nc}.}
\label{fig:maps_ez}
\end{figure*}

\begin{figure}
\centering
\includegraphics{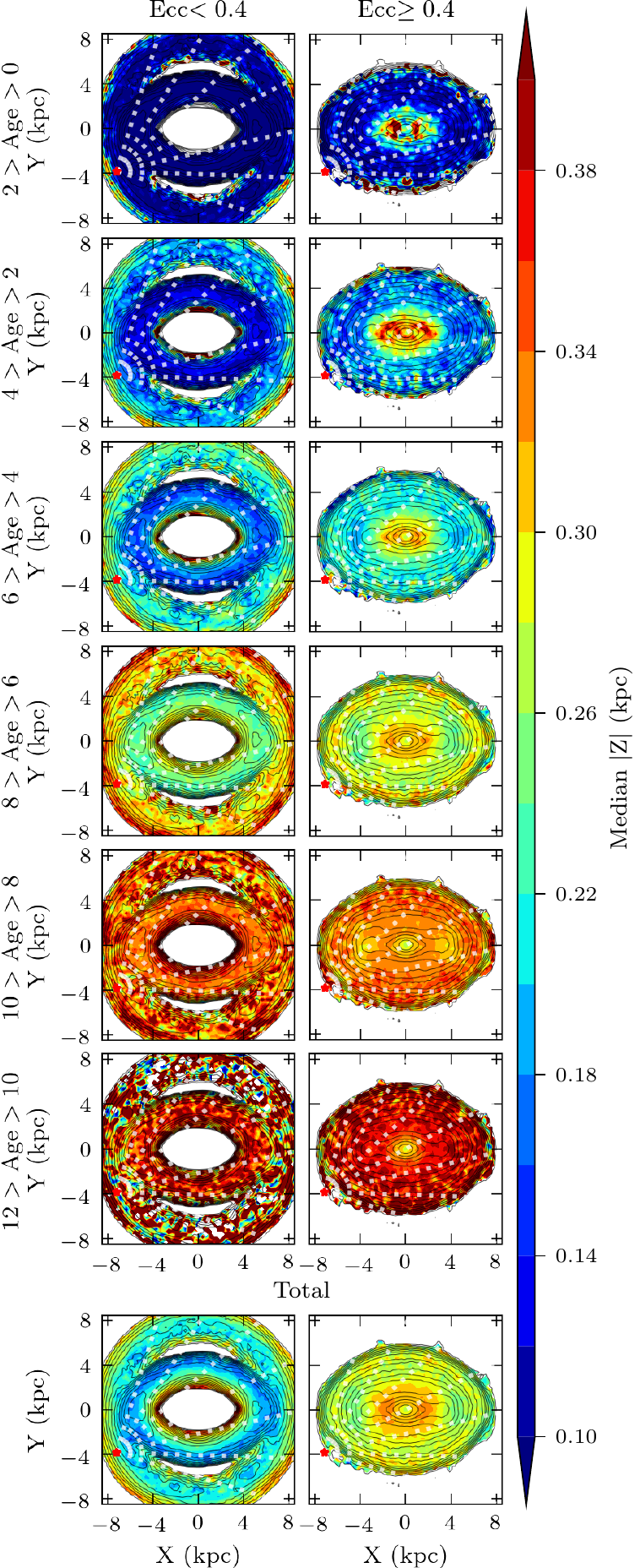}
\caption{Median $|\mathrm{Z}|$ of the stars (orbital time steps) along Y-symmetric $\apogee$ orbits with $|\mathrm{Z}|<0.75$ kpc in the ring and disc (left; $\mathrm{Ecc}<0.4$) and bar (right; $\mathrm{Ecc}\geq0.4$). The top six rows show the median $|\mathrm{Z}|$ of the stars in different age bins, while the last row shows the median $|\mathrm{Z}|$ of the total. The black curves show the orbital density maps built from all disc and ring stars and all bar stars. For other plot details see Fig.\ref{fig:maps_nc}.}
\label{fig:med_ez}
\end{figure}

\begin{figure*}
\centering
\includegraphics{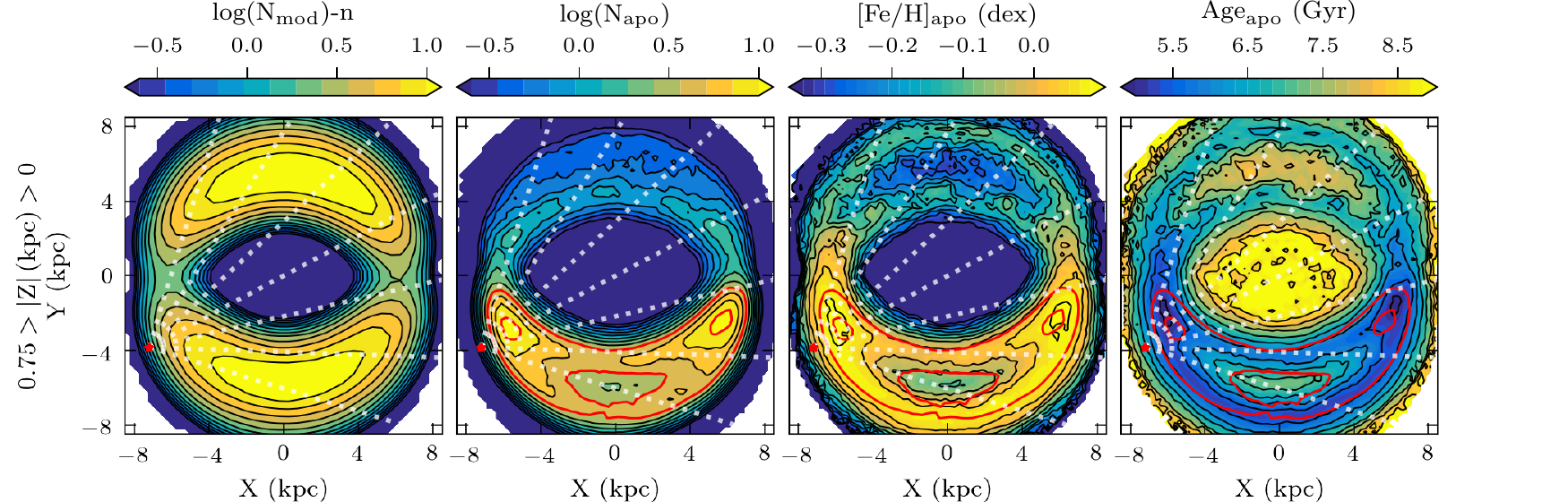}
\caption{Orbital maps of the Y-asymmetric orbits with $|\mathrm{Z}|<0.75$ kpc and $-1<\feh \, (\mathrm{dex})<0.5$. Left two plots: Density maps built by superposing of model (left) and $\apogee$ stellar (right) Y-asymmetric orbits. The model density map has been normalised to a similar scale as the $\apogee$ density map using the factor n. Right two plots: Mean $\feh$ (left) and age (right) maps built from the Y-asymmetric $\apogee$ stellar orbits. The red contours in the $\apogee$ maps show specific density levels to guide the eye. For other plot details see Fig.\ref{fig:maps_nc}.}
\label{fig:maps_antis}
\end{figure*}

\begin{figure*}
\centering
\includegraphics{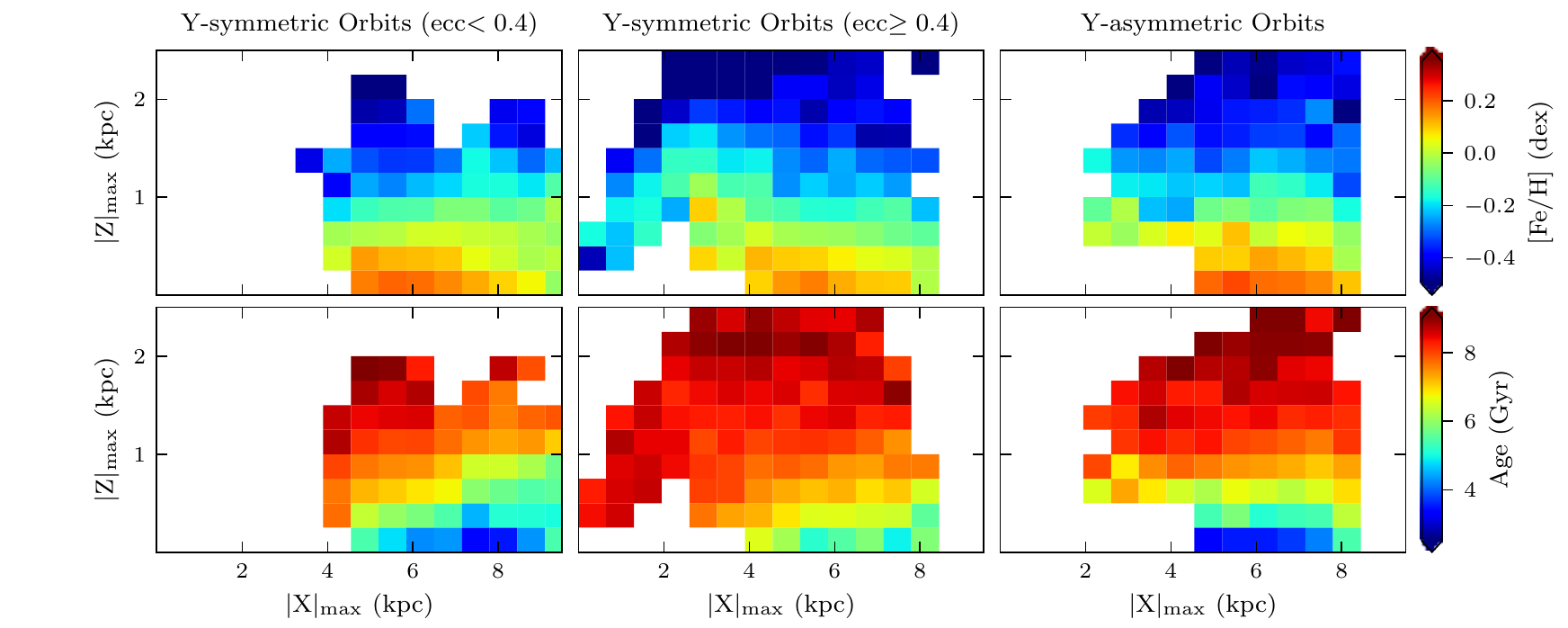}
\caption{$|\mathrm{X}|_{\mathrm{max}}$-$|\mathrm{Z}|_{\mathrm{max}}$ distributions for the Y-symmetric low eccentricity (left), Y-symmetric high eccentricity (middle), and Y-asymmetric (right) $\apogee$ orbits, coloured by mean $\feh$ (top) and AstroNN age (bottom). The white areas are regions with less than ten stars per bin.}
\label{fig:zmax_xmax}
\end{figure*}

In Fig.\ref{fig:maps_nc} we show the orbital maps built for the model particles and the $\apogee$ stars. We restricted the orbits in Fig.\ref{fig:maps_nc} to those that are Y-symmetric (i.e. that spend near-equal amounts of time on both sides of the Y-axis.) This cut mainly removed L4 and L5 Lagrange orbits, centred at $(\mathrm{X}, \mathrm{Y})= (0, {\sim}\pm6)$ kpc, which are strongly affected by the $\apogee$ spatial SF (discussed later). In the left two columns of Fig.\ref{fig:maps_nc}, we reconstructed the density distributions at low and high $|\mathrm{Z}|$ (top and bottom rows, respectively), traced by the model particles (first column) and $\apogee$ stars (second column), by superimposing the orbits of each in the model potential and treating the orbital time steps as individual stars. 

Close to the plane, we see a bar that is roughly 4-5.5 kpc long in both density distributions. As is expected, due to the lack of SF corrections to the $\apogee$ data and to likely residual differences between the model and the MW, the two bars do not completely agree. The $\apogee$ orbital bar has shallower density gradients along its major and minor axes and ansae that are not seen in the model. Interestingly, the $\apogee$ orbital bar appears to have a similar structure to that of the metal-rich stars in the \citet[][]{Portail_2017b} chemodynamical model (same potential as the \citetalias{Portail_dyn_2017} model with $\Omega_b\!=\!40 \, \mathrm{km} \, \mathrm{s}^{-1} \, \mathrm{kpc}^{-1}$), while our model's orbital bar looks very similar to their model's full face on projection (shown in their Figs. 1 and 8). At larger heights from the plane, the bars in the model and from the $\apogee$ orbits both become more elliptical and in better agreement with each other. The X-shape of the bulge is visible in both density maps as two peaks in the density along the bar's major axis. These plots show that the model and the $\apogee$ orbits generally agree on the structure of the bar-bulge region, differing only in part of the detailed substructure.

The right two columns of Fig.\ref{fig:maps_nc} show the bulge's $\feh$ and AstroNN age distributions built from the superposition of the $\apogee$ orbits weighted by their respective densities in each bin. Close to the plane, and in both the $\feh$ and age maps, clear gradients are seen along the major axis of the bar. In $\feh$, there is a positive horizontal gradient (${\sim}0.041 \, \mathrm{dex} \,  \mathrm{kpc}^{-1}$) at low $|\mathrm{Z}|$ along the bar's major axis such that the GC is more $\feh$ poor than the bar ends, illustrating, in terms of the orbits, the horizontal gradient measured by \citet[][their Fig.19]{Wylie_2021}. The age gradient along the major axis is negative such that the stars at the ends of the bar are on average younger than the stars at the GC. 

In both parameter maps, we see an elliptical `inner' ring around the bar at low $|\mathrm{Z}|$ (and weakly at higher $|\mathrm{Z}|$ in $\feh$). This ring is solar in mean $\feh$ and has a mean age of approximately 6 Gyr. At larger heights the stars become, on average, more $\feh$ poor and older due to the vertical gradients in the bulge. In the $\feh$ map, we see two peaks in $\feh$ along the bar's major axis at $\pm 1.5$ kpc. These peaks are due to the X-shape of the boxy/peanut (b/p) bulge \citep[readers are invited to compare also Figs. 16 and 17 in][]{Wylie_2021}.  

To investigate the structures seen in Fig.\ref{fig:maps_nc} further, we divided the orbits into bins of $\feh$ and age, and by low and high cylindrical eccentricity, defined as follows:
\begin{equation}
\mathrm{Ecc} = \frac{\mathrm{max}(\mathrm{R}_{\mathrm{GC}})- \mathrm{min}(\mathrm{R}_{\mathrm{GC}})}{\mathrm{max}(\mathrm{R}_{\mathrm{GC}})+ \mathrm{min}(\mathrm{R}_{\mathrm{GC}})}. 
\end{equation}
While the exact eccentricity cut is somewhat arbitrary, we have chosen it at Ecc$=$0.4 such that it approximately extracts the $\feh$ rich ring that we see in Fig.\ref{fig:maps_nc}. The density distributions of the resulting low ($<0.4$) and high ($\geq0.4$) eccentricity orbits are shown as black contours in Fig.\ref{fig:maps_ez}. The ring and the disc are clearly visible in the low eccentricity distribution. The ring is aligned with the bar and has its highest densities along the bar's major axis. It is composed of orbits whose time-averaged density distributions are roughly elliptical with an average major to minor aspect ratio of ${\sim}0.7$. These orbits are generally confined to small $|\mathrm{Z}|$ (see below) and they have many loops that are more densely spaced near the major axis causing the increase in density there. We also find that, in contrast to the ring, the disc is slightly elongated along the Y-axis.

We see the bar at larger eccentricities. The ansae are no longer present as they are largely caused by the high density regions of the ring along the bar's major axis. The majority of the high eccentricity orbits in the planar bar have $\mathrm{X}_{\mathrm{max}}$ greater than $\mathrm{Y}_{\mathrm{max}}$ by roughly 1 kpc; however, in the mean, the axis ratio is approximately 2:1. We also find that a fraction of the high eccentricity orbits in the planar bar are ring-like. This indicates that a more sophisticated parameter than eccentricity may be required to fully separate the ring from the planar bar. We plan to investigate the orbit structure of the planar bar more thoroughly in the future.

In Fig.\ref{fig:maps_ez} the orbits composing the low and high eccentricity structures are divided into bins of $\feh$ and age, and each bin's fractional contribution to each structure is shown. From these fractional maps, we see that the ring and the disc have a relatively narrow range in $\feh$, mainly solar to slightly super-solar stars, but a wider range in age ($\sim$4-9 Gyr). The ring is slightly more super-solar in $\feh$ than the disc. In contrast to the ring and disc, the inner bulge and bar are mainly composed of older stars with a wide range in $\feh$. Interestingly, both the planar bar and the ring peak in age between 6 and 8 Gyr. The peak in age for the inner bulge is older, between 8 and 10 Gyr.

The young ages of the ring stars indicate that they would be quite concentrated in the Galactic plane. Thus in Fig.\ref{fig:med_ez} we show median $|\mathrm{Z}|$ maps of all stars with $|{Z}|<0.75$ kpc and on Y-symmetric orbits, dividing them into low and high eccentricity as before. The first six rows show the median $|\mathrm{Z}|$ distributions of the stars divided into different age bins, while the last row shows the total. Median $|\mathrm{Z}|$ is a useful measurement as it is indicative of how concentrated the bulk of the stars are in the plane without being overly weighted by outliers. From this plot we see that the distribution of the oldest stars in the ring and disc is ${\sim}3$ thicker than that of the youngest stars. In the total map, the outer ring is thinner than the disc or its inner region, despite the disc being younger  on average (Fig.\ref{fig:maps_ez}). 

In the high eccentricity orbital maps, the inner bulge is thicker than the planar bar at all ages except for the very inner bulge ($<$1kpc), which is thinner than its immediate surroundings. The thick structure persisting to low ages is likely related to the X-shape of the b/p bulge. The planar bar gets thinner with decreasing age. Comparing the low and high eccentricity orbits, we see that the planar bar is of similar to only a slightly greater thickness than the ring in the regions where they spatially overlap and each have a significant number of stars.

We now extend the analysis to Y-asymmetric orbits; in Fig.\ref{fig:maps_antis} we show the density, $\feh$, and age maps of these orbits  for the $\apogee$ stars compared to the density map of the model. In both cases, the majority of the asymmetric orbits are Lagrange orbits trapped in resonance at corotation ($\mathrm{R}_{\mathrm{cr}}\approx6$ kpc). Interestingly, there are very large Lagrange orbits with mean ages of $\sim$6 Gyr and solar to super-solar mean $\feh$ that almost reach the solar radius. While both the model and $\apogee$ agree that these very large Lagrange orbits are populated, Fig.\ref{fig:maps_antis} shows that the total density distributions of all Y-asymmetric orbits disagree; specifically, the $\apogee$ map has a minimum at L4, while the model's map has a maximum at L4. This density minimum in the $\apogee$ map also coincides with a minimum in the $\feh$ map and a maximum in the age map. The density difference as well as the observed $\feh$ and age gradients are likely artefacts of the $\apogee$ spatial SF. The small Lagrange orbits are more distant than the large Lagrange orbits, causing there to be fewer of them in the observed sample. Furthermore, there are gaps in the spatial distribution of the $\apogee$ fields (see the top plot of Fig.\ref{fig:mod_data_helio}) near the locations of L4 and L5. The few stars that are observed there are biased towards larger heights from the plane due the rise of the $\apogee$ sight lines with distance. This causes the small Lagrange orbits in the observed $\apogee$ sample to be biased towards lower $\feh$ and older ages and result in the observed parameter gradients. We see in the next paragraph that these gradients disappear when the stars are restricted in maximum extent from the plane.

In Fig.\ref{fig:zmax_xmax} we divided the $\apogee$ orbits by eccentricity and Y-symmetry and show their $|\mathrm{X}|_{\mathrm{max}}$-$|\mathrm{Z}|_{\mathrm{max}}$ distributions coloured by mean $\feh$ and AstroNN age. For all three orbital types, orbits with $\mathrm{X}_{\mathrm{max}}$>$3$ kpc and confined near the plane are, on average, younger and more $\feh$ rich than those that can reach larger heights. Larger differences appear between the orbital types in their horizontal profiles. For the distributions of the inner ring and disc orbits (first column), the horizontal in-plane $\feh$ gradient is slightly negative, such that the ring is more $\feh$ rich than the disc. The age profile differs with the mean age decreasing with increasing $|\mathrm{X}_{\mathrm{max}}|$ until $7-8.5$ kpc after which it increases again. 

Stars on the more eccentric bar and bulge orbits (middle column) have roughly a constant mean $\feh$ and age out to $7$ kpc, from whereon both $\feh$ and age decrease. The gradients of the eccentric orbits change for $\mathrm{X}_{\mathrm{max}}$<$3$ kpc. There the stars also become more $\feh$ rich towards the plane until 0.5-0.75 kpc, below which they start to become more $\feh$ poor again. Thus the inner bulge is more $\feh$ poor than the bar. We also again see that the inner bulge is older than the bar. These trends indicate that the gradients we see in Figs.\ref{fig:maps_nc} and \ref{fig:maps_ez} along the bar's major axis are real and not the result of the $\apogee$ spatial SF. 

In the last column which shows the distributions of the Lagrange orbits, the stars are consistent with having a constant mean $\feh$ and a constant or slightly younger mean age with decreasing extent in $\mathrm{X}_{\mathrm{max}}$. This figure confirms that the gradients in Fig.\ref{fig:maps_antis} are the result of the $\apogee$ spatial SF. The trends seen in this figure do not clearly support the trend measured by \citet{Chiba_2021}, but more data close to the L4 point would be needed to be sure.

%%%%%%%%%%%%%%%%%%%%%%%%%%%%%%%%%%%%%%%%%%%%%%%%%%%%%%%%%%%%%%%%%%%%%%%%
\section{Discussion and conclusions}\label{conclusions}
In order to understand better the structure of the MW's bar and the horizontal metallicity and age gradients seen in the $\apogee$ data, we integrated the orbits of $32,536$ $\apogee$ stars in the gravitational potential of a realistic dynamical model of the inner Galaxy which was fitted to star counts and line-of-sight velocities (P17; see Sect.\ref{methods}). We find that stellar orbits in the outer parts of the planar bar trace a radially thick, vertically thin, elongated, stellar ring of low-eccentricity orbits, with a face-on width of $\sim$2-3 kpc and an axis ratio of $\sim$0.7. This ring is seen outwards of a major axis distance of X$\sim$4 kpc, while the remaining planar bar reaches X$\sim$4.5 kpc. Both components together make up a shallow outer long bar similar to that seen in the dynamical model. While there are quantitative variations with the bar model, this basic structure is robust for similar dynamical models with pattern speeds in the range 35-$40 \, \mathrm{km} \, \mathrm{s}^{-1} \, \mathrm{kpc}^{-1}$ favoured by recent studies, as mentioned in Sec. \ref{intro}.

Colouring the orbits by their ASPCAP $\feh$ and AstroNN ages, we find that the stellar ring is dominated by stars with solar to slightly super-solar metallicities ($\feh=0.125\pm0.25$ dex), and ages in the range $4-$9 Gyr with a peak at 7 Gyr and a long tail towards younger ages. Ring stars with young ages are highly concentrated towards the plane. The eccentric planar bar is more metal-poor with a wider range of metallicities and is more concentrated in age around $6-$9 Gyr with a peak at 8 Gyr. The bar ends are, on average, younger and more $\feh$ rich than the b/p bulge, illustrating the horizontal gradients seen previously. The bulge's X-shape is clearly visible in $\feh$ and density.

\citet{Sevenster_1999} previously suggested that the 3 kpc arm of the MW could actually be an inner ring and, more recently, from their study of the gas dynamics of the inner MW, \citet{Li_2021} suggest that the 3~kpc arm along with the Norma arm and the bar-spiral interfaces compose an inner gas ring. Furthermore, \citet{Kormendy_2019} find that two MW analogues, NGC 4565 and NGC 5746, host inner rings.

In Sbc galaxies, such as the MW, a close connection between inner rings and spiral arms is common. \citet{Buta_2015} find that of the $45\%$ of the S4G Sbc galaxies that host inner rings, most tend to be ring spirals as opposed to pure rings or ring lenses, which are more common in early types. The model from which we generated the potential does not include spiral arms, making it impossible for us to see a connection between the inner stellar ring and the spiral arms that might exist in the MW. To understand better whether this structure in the MW is a ring or ring-spiral, the model must be augmented to include a realistic spiral arm potential.

The in-plane thickness of the inner ring is somewhat surprising as inner rings are generally thinner. Passive rings, which are rings that are no longer forming stars, can be thicker; however, they are only seen in earlier type galaxies \citep[$\geq$ Sab,][]{comeron_2013}. It is possible that the MW also hosts a thin star-forming ring, but due to observing in the infrared, we can only see its older part which has spread out over time \citep[as may be the case for NGC 7702,][]{Buta_1991}. Alternatively, the ring could actually be a lens such that the MW hosts both a lens and a bar. Lenses are characterised by a sharp edge encircling a shallow brightness gradient. However, further work is required to better understand the sharpness of the transition from the ring to the surrounding disc, and as discussed earlier, a lens would be unusual for a galaxy as late-type as the MW \citep{Buta_2015}.

The question of whether the stellar inner ring is a separate structure from the long bar is not easily answered and may depend on how both structures are defined. The figures above show that the metallicity distribution function of the ring is shifted to higher metallicities than that of the bar, with a strong peak at super-solar metallicities, and the age distribution function to $\sim$1 Gyr lower mean age, with a long tail towards younger ages. This leads to clear outward metallicity and age gradients in the bar region. In the orbit distribution, the ring has its highest density on lower eccentricity orbits close to corotation. Increasing the threshold in eccentricity to 0.5 results in a slightly thicker ring and shorter bar. However, some high-eccentricity orbits in the bar are still ring-like in appearance, there are high-metallicity and intermediate-age orbits in the bar as well as old stars on resonant ring orbits, and both components together make up a shallow outer long bar in the rotating potential. That is, in many variables, the distributions overlap, suggesting that the ring and bar are not easily separable, discrete components. The highly flattened, young components in both the ring and the bar were probably formed from gas and added to the old metal-poor central bulge and bar recently. The late addition of stars causes the overall potential and orbit structure of the system to evolve, further complicating matters. In all, this warrants a more careful study of the bar-ring connection.

The spatial transition between the ring and planar bar occurs at around $\sim4$ kpc. Comparing this scale with the transition reported in \citet{Wegg_2015} between their `thin' and `superthin' bars suggests that the superthin bar, which dominates outside $\mathrm{X}=4.$5kpc, might be identifiable with the young and thin part of the stellar ring in the $\apogee$ orbits. This also requires further study.
 
In the orbit maps, we also find very large Lagrange orbits that almost reach the solar radius and have a solar to super-solar mean metallicity and intermediate age (${\sim} 6$ Gyr). The small Lagrange orbits are not well sampled by $\apogee$ due to its spatial SF, but some can be found by dividing orbits in terms of their $|\mathrm{X}|$ and $|\mathrm{Z}|$ maxima. These suggest that stars closer to the core of the resonance have similar [Fe/H] and ages as the outer Lagrange orbits. This does not clearly support the trend measured by \citet{Chiba_2021}, but further investigation with more data close to the L4 point is needed.

In the two barred galaxy models with a metal-rich inner ring in the Auriga cosmological simulations, the ring continues to form stars from gas driven inside corotation after the formation of the bar while the star formation in the bar drops more rapidly \citep{Fragkoudi_2020}. They argue that the time when the star formation in the bar starts quenching may be used to estimate a lower limit to the age of the bar. In the AstroNN age distributions for the $\apogee$ stars, the peak for the fraction of stars in the inner ring occurs at age $\simeq 7$ Gyr; whereas for the stars in the remaining planar bar outside the bulge, it occurs at $\simeq 8$ Gyr. After age $\simeq 7$ Gyr, the fraction of newly formed bar stars drops more rapidly than for the ring stars; therefore, we used this to estimate that the Galactic bar formed at least 7 Gyr ago.

In conclusion, by studying the orbits of $\apogee$ stars in the gravitational potential of a dynamical bar model for the MW with a slow $\Omega_b$ as measured recently, we find that (i) the MW hosts a radially thick, vertically thin elongated inner ring; (ii) the ring is dominated by stars with solar to super-solar metallicities and with ages between 4 and 9 Gyr; and (iii) the ring explains the steep horizontal metallicity gradient along the Galactic bar's major axis. From the distribution of ages in the ring and planar bar, we estimate a lower limit of 7 Gyr to the time since the formation of the bar.

\begin{acknowledgements}\\
We thank Grzegorz Gajda for helpful discussions and Jo Bovy and Henry Leung for advice on the AstroNN distances. We thank the $\apogee$ team for making the $\apogee$ DR16 catalogue publicly available. Funding for the Sloan Digital Sky Survey IV has been provided by the Alfred P. Sloan Foundation, the U.S. Department of Energy Office of Science, and the Participating Institutions. This work has made use of data from the European Space Agency (ESA) mission {\it Gaia} (\url{https://www.cosmos.esa.int/gaia}), processed by the {\it Gaia} Data Processing and Analysis Consortium (DPAC, \url{https://www.cosmos.esa.int/web/gaia/dpac/consortium}). Funding for the DPAC has been provided by national institutions, in particular the institutions participating in the {\it Gaia} Multilateral Agreement.
\end{acknowledgements}

\section*{ORCID iDs}
\noindent S.M. Wylie \orcid{0000-0001-9116-6767} \href{https://orcid.org/0000-0001-9116-6767}{https://orcid.org/0000-0001-9116-6767}\\
\noindent J.P. Clarke \orcid{0000-0002-2243-178X} \href{https://orcid.org/0000-0002-2243-178X}{https://orcid.org/0000-0002-2243-178X}\\
\noindent O.E. Gerhard \orcid{0000-0003-3333-0033} \href{https://orcid.org/0000-0003-3333-0033}{https://orcid.org/0000-0003-3333-0033}

\bibliographystyle{aa}
\bibliography{mknbib}

\end{document}